REVTEX 3.0

# Experimental Evidence for a Coulomb Gap in Two Dimensions

Whitney Mason, S. V. Kravchenko, G. E. Bowker*, and J. E. Furneaux
*Laboratory for Electronic Properties of Materials and Department of Physics and Astronomy, University of Oklahoma, Norman, Oklahoma 73019*

We have studied the resistivity of a two-dimensional electron system in silicon in the temperature range 200 mK < $T$ < 7.5 K at zero magnetic field at low electron densities, when the electron system is in the insulating regime. Our results show that at an intermediate temperature range, $\rho = \rho_0 \exp[(T_0/T)^{\frac{1}{2}}]$ for at least four orders of magnitude up to $3\times10^9$ Ω. This behavior is consistent with the existence of a Coulomb gap. Near the metal/insulator transition, the prefactor was found to be $\rho_0 \approx h/e^2$, and resistivity scales with temperature. For very low electron densities, $n_s$, the prefactor diminishes with diminishing $n_s$. A comparison with the theory shows that a specific set of conditions are necessary to observe the behavior of resistivity consistent with the existence of the Coulomb gap.

At sufficiently low temperatures ($T$), in disordered systems such as semiconductors, transport occurs by phonon-assisted tunneling to states nearby in energy. The tunneling distance to a state within $k_BT$ of the Fermi energy ($E_F$) increases with decreasing temperature (here $k_B$ is the Boltzmann constant). This transport process has been labeled variable-range hopping (VRH) and is characterized by resistivity of the form

$$\rho(T) = \rho_0 \exp{(T_0/T)^x}, \quad (1)$$

where $T_0$ is some characteristic temperature. Mott [1] derived this law by assuming a constant density of states (DOS) at the Fermi energy and found in two dimensions that $x = \frac{1}{3}$. This is a single-particle picture which ignores the Coulomb interaction. Efros and Shklovskii [2] have argued that the Coulomb interaction between localized electrons creates a gap, the so-called "Coulomb gap", in the density of states near the Fermi energy. This is manifested by a resistivity of the form of Eq. (1) with $x = \frac{1}{2}$, which is universal for both two- and three-dimensional (2D and 3D) electronic systems.

In 1986, Timp, Fowler, Hartstein, and Butcher (TFHB) [3] examined the conductivity as a function of temperature and electric field in sodium-doped silicon metal-oxide-semiconductor field-effect transistors (MOSFET's) and found no evidence of the Coulomb gap. More recently, Coulomb gap behavior has been observed in relatively low-mobility GaAs/AlGaAs heterostructures [4,5]. In this paper, we report experimental studies of the temperature dependence of resistivity of *high-mobility* silicon MOSFET's. Our experimental data follows the form of Eq. (1) with $x = \frac{1}{2}$ for a range of parameters such as temperature and 2D electron density $n_s$. This behavior is consistent with the existence of the Coulomb gap. We believe differences between our results and those of TFHB are due to differences between parameter spaces examined in each case.

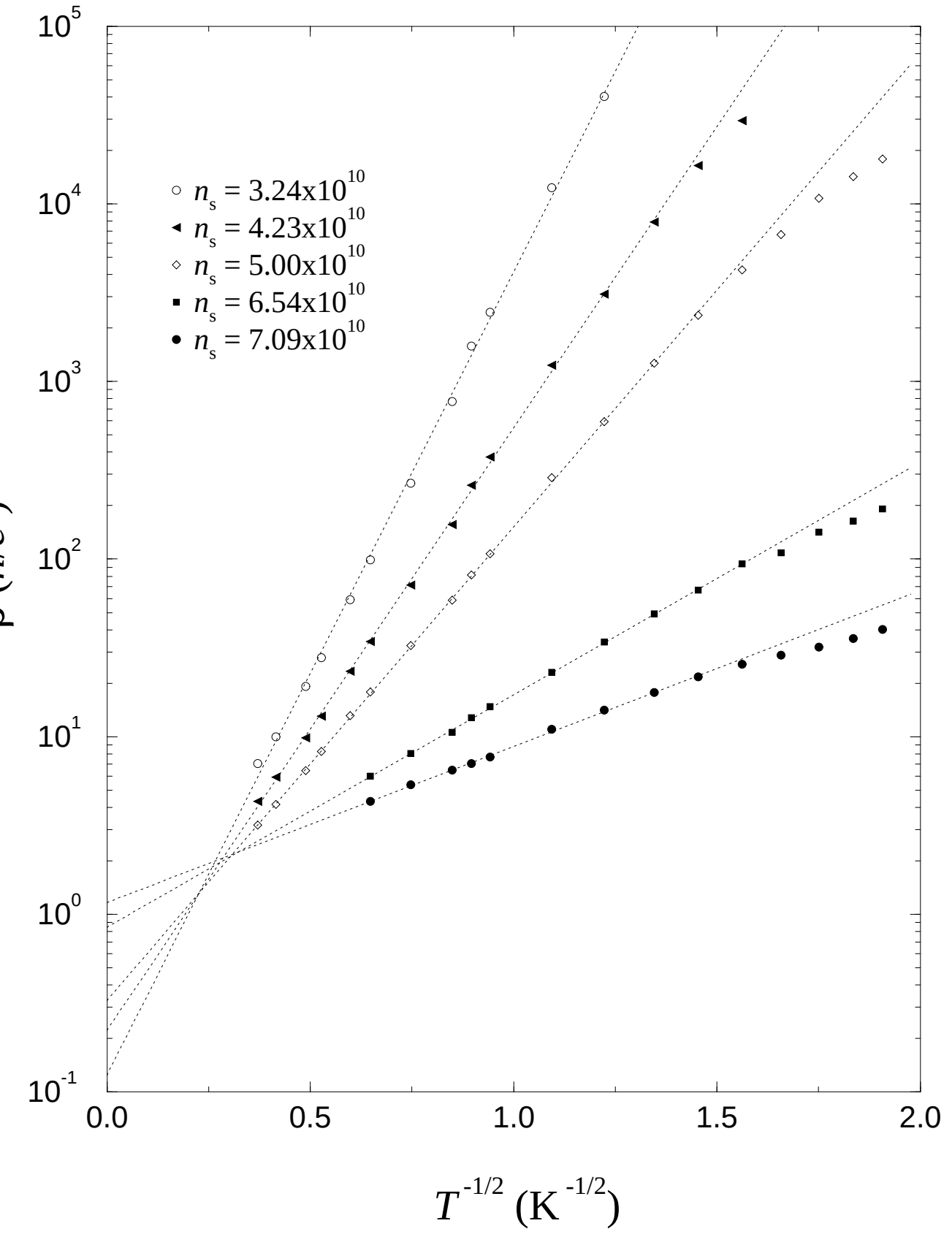

FIG. 1. Resistivity vs $T^{-\frac{1}{2}}$ for sample Si-1 in the insulating regime for various electron densities.

The samples used for the studies reported here were 2D electron systems (2DES) in silicon MOSFET's with maximum mobility, $\mu$, of $3.0\times10^4$ cm$^2$/Vs (sample Si-1) and $3.5\times10^4$ cm$^2$/Vs (Si-2). The gate oxide thickness is 200 nm. Electron density was changed by changing the voltage, $V_g$, between the 2DES and the gate. The resistance was measured with a four-terminal dc technique using cold amplifiers with a very high input impedance ($> 10^{14}$ Ω). The data discussed here is for electron densities where the samples exhibited strongly insulating behavior. In this case, the current-voltage ($I - V$) characteristics are strongly nonlinear [6]. To get a "true" resistivity, we measured $V(I)$ for each $n_s$ and $T$; then the values of resistivity corresponding to $I \rightarrow 0$ were obtained as $(dV/dI)|_{I=0}$ divided by the number of squares. The measurements were taken at temperatures ranging from 200 mK to 10K at $B = 0$.

TABLE I. Values of the best fit for $x$ in Eq. (1) for sample Si-1 and the parameters discussed in the text for various electron densities.

| $n_s(10^{10} cm^{-2})$ | $x$ | $T_0$(K) | $\xi$(nm) |
|---|---|---|---|
| 3.24 | 0.494±0.007 | 108 | 120 |
| 4.23 | 0.493±0.005 | 61 | 200 |
| 5.00 | 0.500±0.002 | 38 | 330 |
| 6.54 | 0.500±0.004 | 9.1 | 1400 |
| 7.09 | 0.493±0.006 | 4.1 | 3000 |

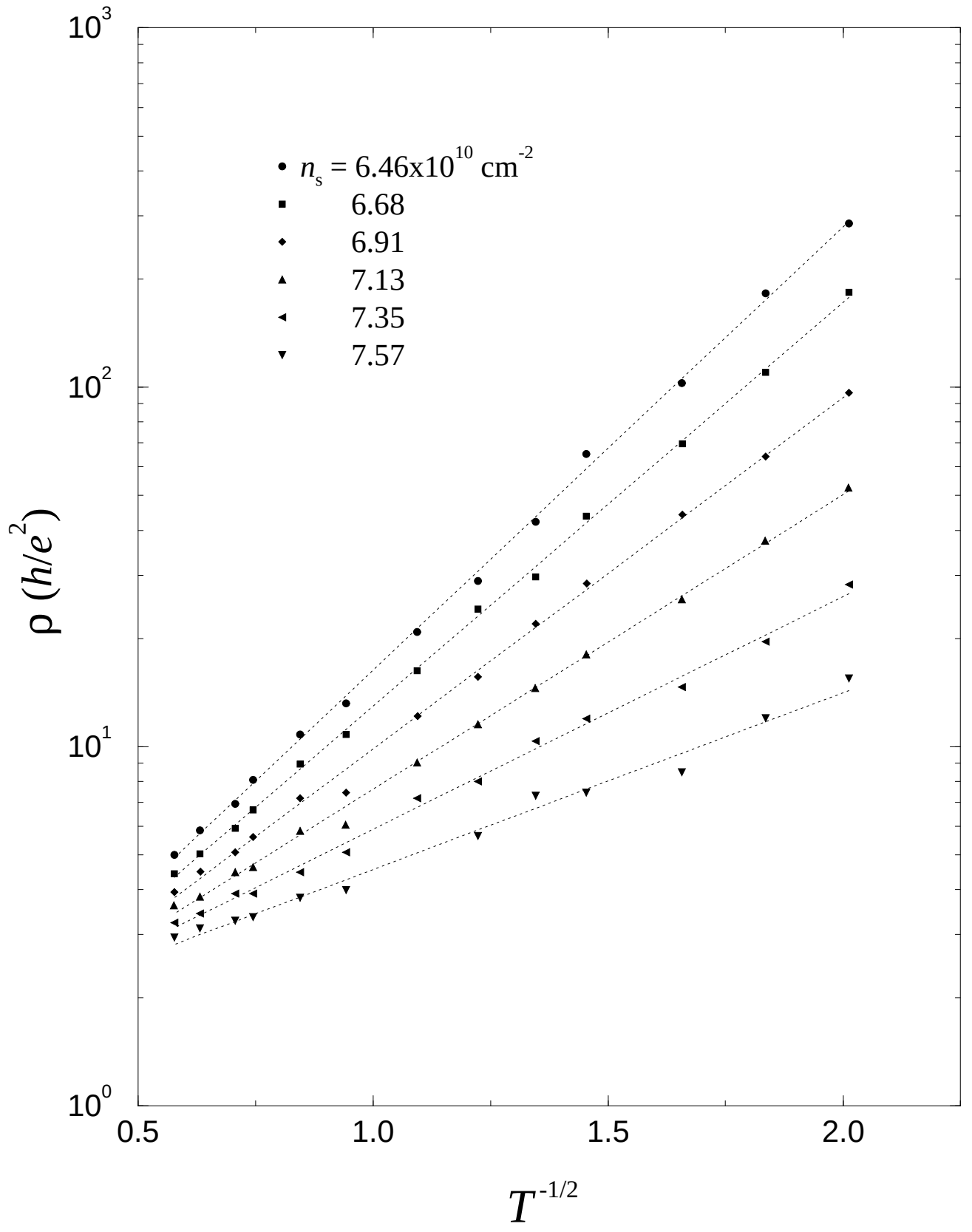

FIG. 2. Resistivity vs $T^{-\frac{1}{2}}$ for sample Si-2 for a narrow interval of $n_s$ near the metal/insulator transition.

Figure 1 shows $\rho$ vs $T^{-\frac{1}{2}}$ for Si-1 for several electron densities [7] ($\rho$ is given in units of $h/e^2$). In order to determine the value of $x$ in Eq. (1), we fit our data to $\exp{(T^x)}$ assuming a temperature-independent prefactor, $\rho_0$, and found that for $T \gtrsim 400$mK, the best fit of all the data gave values of $x$ very close to 0.5. The actual values of $x$ for various electron densities are shown in Table I. One can see that the Coulomb gap behavior exists for up to four orders of magnitude in resistivity. At low temperature weaker T dependence is evident, particularly at higher electron density. Notice that the data are consistent with a *temperature-independent prefactor*, unlike the data presented in Ref. [5]. However, it can be seen that the prefactor is $n_s$-dependent: the lower $n_s$, the lower the prefactor.

Figure 2 shows $\rho$ as a function of $T^{-\frac{1}{2}}$ for sample Si-2 with slightly higher mobility for a narrow interval on electron densities close to the metal/insulator transition [8]. These curves show Coulomb gap behavior for about two orders of magnitude. Perhaps the most impressive part of this data is that all curves presented in Fig. 2 can be made to collapse onto one curve by scaling them along the $T$-axis. The resulting curve is shown in Fig. 3. Note that the curve extrapolates to the value of $h/e^2$, the quantum of resistivity, though the curve deviates from the $T^{-\frac{1}{2}}$ behavior when $T$ becomes comparable with $T_0$; however, the data scale even then. Note that in the quantum Hall effect regime, where the Coulomb gap behavior has also been observed [9,10], the prefactor $\sigma_0$ near the critical point was found to have a form $e^2/h \cdot f(T/T_0)$ with $f \sim 1$ at $T \sim T_0$. The nature of this universal prefactor is unclear because the theory of phonon-assistant hopping gives much lower value [11]. Aleiner, Polyakov, and Shklovskii suggested [11] that this discrepancy might be resolved if the hopping was assisted by electron-electron scattering rather than by electron-phonon one.

In the interacting picture [2], $T_0$ in Eq. 1 is equal to

$$T_0 = \frac{Ce^2}{k_B \epsilon \xi}. \quad (2)$$

Here $e$ is the electron charge, $\epsilon$ is the dielectric constant, here given an average value of 8, $\xi$ is the localization length, and $C$ is a constant which has been found [13] to be about 6.2. Therefore, it is possible to get the localization length from the slope of $\log \rho$ vs $T^{-\frac{1}{2}}$ curves. Table I gives values of the localization length for representative electron densities. Because the gate is a good metal, the effective distance to the gate, $d$, plays the role of a screening radius [12]. In other words, the Coulomb interaction between charges separated by more than $\sim 2d$ will be effectively screened by the metal gate. Thus the

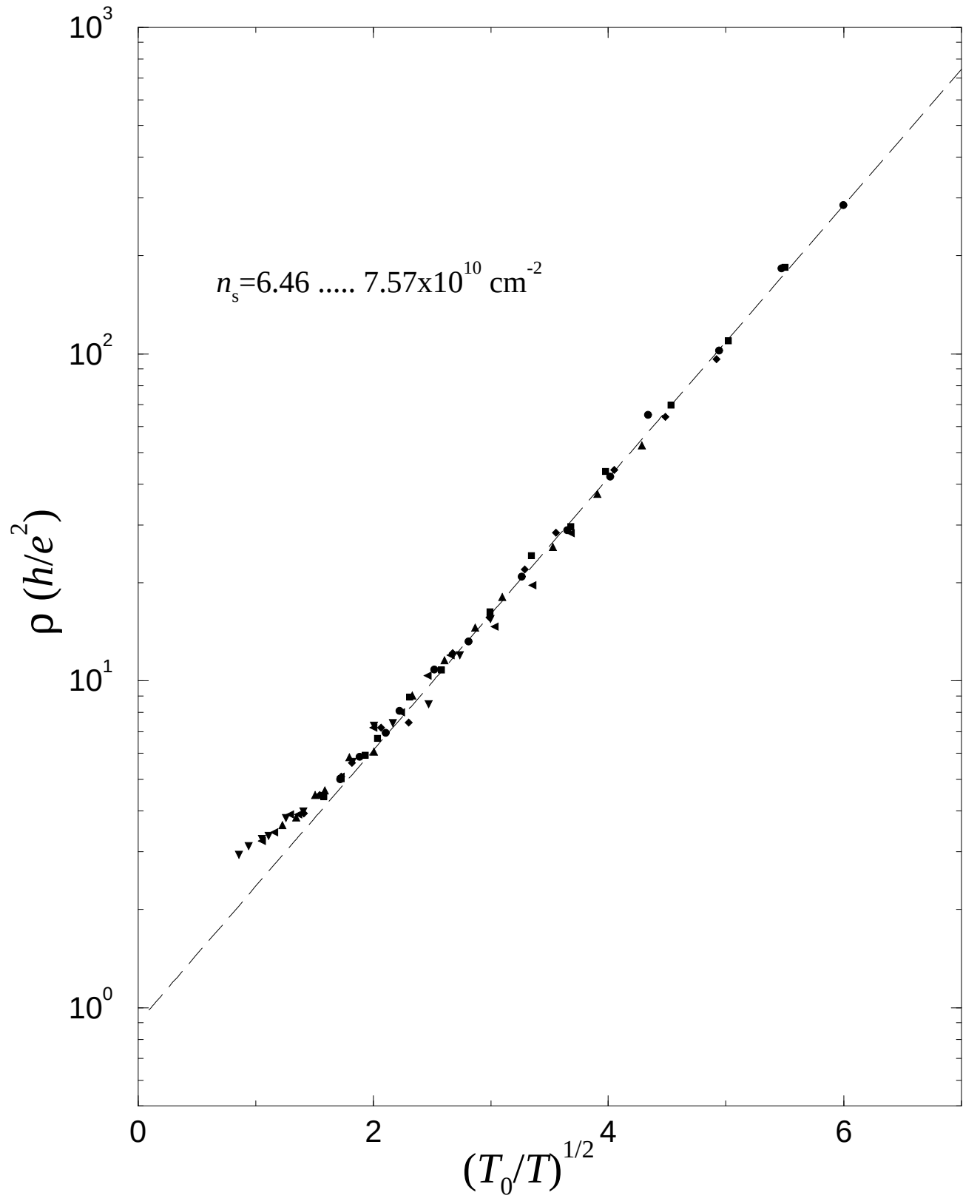

FIG. 3. Resistivity as a function of $(T_0/T)^{\frac{1}{2}}$ for Si-2.

Coulomb gap should exist only when the hopping length,

$$r_h \sim \frac{1}{4}\,\xi\,(T_0/T)^{1/2}, \tag{3}$$

is shorter than $2d$, while in the opposite case, when $r_h$ is a few times $d$, temperature dependence of resistivity should obey the Mott's law, $\rho(T) \propto \exp(T_0/T)^{1/3}$. If $\xi \ll d$, the crossover between the two regimes should happen at the temperature calculated by Aleiner and Shklovskii [12] to be

$$T \sim 0.013\,\frac{e^2\xi}{k_B\epsilon d^2}. \tag{4}$$

The net effect of the condition for Coulomb gap formation that $2d > \xi$ is to place an upper bound on the $n_s$ where the Coulomb gap can form. However, for the lowest curve in Fig. 1 ($T_0 = 4.1$ K), we see a Coulomb gap behavior down to $T \approx 400$ mK which, according to Eq. 3, corresponds to $r_h \approx 2400$ nm, i.e., 6 times larger than $2d$. Perhaps the coefficient $C$ in Eq. (2) has a smaller value than 6.2 which we used calculating $\xi$ and $r_h$ from the data for $T_0$; however, the nature of a smaller $C$ is not known at present. Another reason, which seems unlikely, could be that the dielectric constant, $\epsilon$, is bigger than the lattice value. It is worth noting that a quantitative agreement between experiment [5] and theory [12] was achieved after assuming that the dielectric constant is 4.8 times bigger than the lattice value.

The screening of electron-electron interactions by the gate leads to reconciliation of the results of TFHB. They found that when they could fit their data to $T^{-\frac{1}{2}}$, the data fit over some limited temperature range, but the behavior deviated to follow $T^{-\frac{1}{3}}$ as the temperature was lowered. This supports the idea of Aleiner and Shklovskii that the Coulomb gap behavior occurs in some intermediate temperature range, and that as the temperature is lowered, so that the hopping distance becomes $r_h \gg 2d$, the resistivity (conductivity) must eventually follow Mott's law.

In conclusion, the data presented here is consistent with the existence of the Coulomb gap in two dimensions. The resistivity was found to follow the Coulomb gap law, $\rho = \rho_0 \exp[(T_0/T)^{-\frac{1}{2}}]$ with temperature-independent prefactor for four orders of magnitude. In some range of $n_s$ near the metal/insulator transition, the prefactor is close to the quantum of resistivity, $h/e^2$, and resistivity scales with temperature. At lower $n_s$, the prefactor was found to decrease as $n_s$ is decreased.

We acknowledge useful discussions with B. I. Shklovskii. This work was supported by grants DMR 89-22222 and Oklahoma EPSCoR via LEPM from the National Science Foundation.

* Permanent address: Department of Physics and Astronomy, Bowdoin College, Brunswick, Maine 04011.